\documentclass[12pt]{article}
\usepackage{graphicx}
\usepackage{cite}
\def\be{\begin{equation}}
\def\ee{\end{equation}}
\def\ba{\begin{eqnarray}}
\def\ea{\end{eqnarray}}
\def\la{~\mbox{\raisebox{-.6ex}{$\stackrel{<}{\sim}$}}~}
\def\ga{~\mbox{\raisebox{-.6ex}{$\stackrel{>}{\sim}$}}~}
\def\bq{\begin{quote}}
\def\eq{\end{quote}}

 at 10truept

\newcommand{\beq}{\begin{equation}}
\newcommand{\eeq}{\end{equation}}
\newcommand{\beqa}{\begin{eqnarray}}
\newcommand{\eeqa}{\end{eqnarray}}

\def\la{~\mbox{\raisebox{-.6ex}{$\stackrel{<}{\sim}$}}~}
\def\ga{~\mbox{\raisebox{-.6ex}{$\stackrel{>}{\sim}$}}~}

\def\ga{~\mbox{\raisebox{-.6ex}{$\stackrel{>}{\sim}$}}~}
\def\ltap{\ \raise.3ex\hbox{$<$\kern-.75em\lower1ex\hbox{$\sim$}}\ }
\def\gtap{\ \raise.3ex\hbox{$>$\kern-.75em\lower1ex\hbox{$\sim$}}\ }
\def\gl{\ \raise.5ex\hbox{$>$}\kern-.8em\lower.5ex\hbox{$<$}\ }
\def\roughly#1{\raise.3ex\hbox{$#1$\kern-.75em\lower1ex\hbox{$\sim$}}}

\begin{document}

\thispagestyle{empty}
\begin{flushright}
{\tt astro-ph/0507148}\\
\end{flushright}

\vskip2cm
\begin{center}
{\LARGE{\bf The Accelerated Acceleration of the Universe}} \\
\vskip2cm {\large Csaba Cs\'aki$^{a,}$\footnote{\tt
csaki@lepp.cornell.edu}, Nemanja Kaloper$^{b,}$\footnote{\tt
kaloper@physics.ucdavis.edu} and John Terning$^{b,}$\footnote{\tt
terning@physics.ucdavis.edu}}\\

\vspace{.5cm}

$^a${\em Newman Laboratory of Elementary Particle Physics, Cornell
University, }\\
{\em Ithaca, NY 14853}\\

\vskip 0.15in

$^b${\em Department of Physics, University of California,}\\
{\em Davis, CA 95616}\\

\vskip 0.15in

\vskip 0.1in \vskip 0.1in
\end{center}
\vskip .25in
We present a  simple mechanism which can mimic dark energy with an
equation of state $w < -1$ as deduced from the supernova data. We
imagine that the universe is accelerating under the control of a
quintessence field, which is moving {\it up} a very gently sloping
potential. As a result, the potential energy and hence the
acceleration increases at lower redshifts.  Fitting this behavior
with a dark energy model with constant $w$ would require $w<-1$.
In fact we find that the choice of parameters which improves the
fit to the SNe mimics $w = -1.4$ at low redshifts. Running  up the
potential in fact provides the best fit to the SN data for a
generic quintessence model. However, unlike models with phantoms,
our model does not have negative energies or negative norm states.
Future searches for supernovae at low redshifts $0.1 < z < 0.5$
and at high redshifts $z>1$ may be a useful probe of our proposal.


\vfill \setcounter{page}{0} \setcounter{footnote}{0}
\newpage

\setcounter{equation}{0} \setcounter{footnote}{0}

A wide range of observational evidence indicates that our universe
may be accelerating \cite{sne,Riess,cmb,krtur}. If we assume that
long-range gravity obeys Einstein's General Relativity (GR), this
suggests that most of our universe is in some form of smooth dark
energy, which can comprise  $\sim 70\%$ of the critical energy
density. In order to drive cosmic acceleration this dark energy
must have negative pressure, which should satisfy $w = p/\rho \la
-2/3$ to fit the observations \cite{eqst,cmblss}. Benchmark models
of dark energy are a cosmological constant or a time-dependent
quintessence field \cite{linq,q,ams,coincidence}, which must be
finely tuned to fit the data \cite{weinberg,carroll}. There are
but a few examples of quintessence which are natural from the
point of view of the 4D effective field theory, where quintessence
is a pseudo-scalar Goldstone boson \cite{axq}, with a radiatively
stable mass and naturally weak couplings to the visible matter.
There are also models where dark energy is in the form of a
network of domain walls, whose coarse-grained dynamics is
described by an equation of state $w = -2/3$ \cite{walls}.

One of the greatest challenges in modern cosmology is to discern
the nature of dark energy. To this end, currently the most
sensitive probe of dark energy are the Type Ia supernovae
\cite{sne}.  However, the results of supernova observations cannot
completely constrain the dark energy equation of state yet. First
of all, the supernova data give us the Hubble diagram for
luminosity distance versus redshift, which is related to the dark
energy equation of state parameter $w$ through a double integral
\cite{wellalb,maor,starob}. Therefore there is a great deal of
degeneracy between dark energy models with different, and
variable, equations of state parameters $w$. Further, there may be
additional sources of supernova dimming, which do not imply cosmic
acceleration, as for example the gray dust of \cite{aguirre}.
Another example is the photon-axion conversion which we have
proposed in \cite{ckt,exo}. In this case, the extragalactic
magnetic fields may catalyze the conversion of photons into very
light axions which renders the supernovae dimmer. While the CMB
and large-scale structure alone support the existence of dark
energy \cite{cmblss}, they do not yet imply stringent bounds on
the equation of state, and the ones obtained from supernovae
observations can be significantly relaxed with photon-axion
conversion \cite{ckt,exo}. Various aspects of the photon-axion
mechanism have been considered in
\cite{ckt,exo,ckpt,others,gumor,lknox,Mirizzi,fairbairn}.

Exploiting the limitations of the current data sets, researchers
have even argued that more exotic models of dark energy, yielding
$w < -1$ are allowed \cite{caldwell,ckw,mmot}. The simplest models
of this kind stretch the philosophy of quintessence scalars by
postulating that the scalar is a {\it phantom}: i.e. a ghost,
having negative kinetic terms \cite{caldwell,ckw,mmot,gibbons}.
Phantoms are in fact scaled down models of super-exponential
inflation, invented by Pollock in 1985 \cite{pollock}. In the
realm of GR, they inevitably violate the dominant energy
condition, $|p|\le \rho$, yielding energy density which {\it
increases} with the expansion of the universe and produces a
future singularity. They are also plagued with instabilities even
without GR: they do not have a stable ground state
\cite{gibbons,cht,wise}, have classical runaway modes
\cite{gibbons}, and violate the lore of effective field theory
\cite{cht}. The debate of whether $w$ may have dipped below $-1$
at low redshifts still continues
\cite{starobinsky,huterer,bergstrom}, and while it is not at all
clear from the data that the inferences on $w$ being below $-1$
are reliable \cite{maor,wang}, there has been a lot of exploration
of phantoms \cite{phantasms}.

In our opinion, probing for $w<-1$ (see e.g. \cite{kabri}) would
be {\it considerably} more motivated if less occult
methods for predicting $w<-1$ were easier to
come by. Chasing phantoms is interesting, but one might prefer to
have better prospects for discovering a new twist to dark energy
based on $w<-1$. Motivated by this line of thought, we have
already exorcised the phantom from $w<-1$ once \cite{exo}, having
shown that photon-axion conversion combined with cosmic
acceleration driven by a cosmological constant can easily fool an
observer into thinking that $w<-1$, in fact faking $w$ as low as
$-1.5$. Another, less efficient and more fine tuned, method
to fake $w < -1$ without ghosts could be to weaken gravity in the
far infrared \cite{cdft,lue}. Thus at least in principle it is
possible to have the dark energy equation of state parameter $w$
{\it appear} to be more negative than $-1$ without any phantasms.

\begin{figure}[htb]
\begin{center}
\includegraphics[width=0.7\hsize]{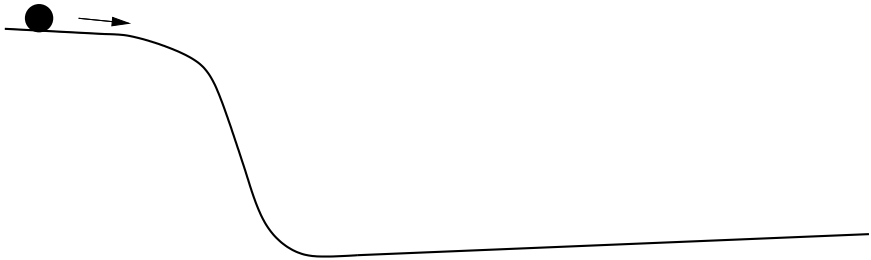}
\end{center}
\caption{Sketch of the form of the potential including an example of
the possible non-linear behavior
before the onset of the linear regime.}
\label{fig:potential}
\end{figure}

In this note, we present yet another method to mimic $w<-1$. It is
very simple. It involves a quintessence field going {\it up} the
potential slope. Imagine that the quintessence potential is
asymmetric around some minimum, with a curvature which may change
by ${\cal O}(\phi)$ contributions as the field passes through the
minimum. The field may end up initially frozen by Hubble friction
on a steeper side of the potential well. Then as the universe
cools, the field is eventually released as $H$ decreases below the
effective mass, and starts to roll down the steeper slope picking
up speed (see Fig. \ref{fig:potential}). We imagine that this
occurs some time around $z=2-3$. As the field picks up speed the
kinetic energy is converted into potential energy and the universe
will eventually be dominated by dark energy. Because the
quintessence field is rolling up the  potential, the cosmic
acceleration increases at lower redshifts, lifting the Hubble
diagram curve up at low redshifts. The quicker the increase, the
higher the lift. Reproducing such behavior in a model of dark
energy with a constant $w$ requires $w<-1$. Because the effect is
embedded in the metric relations in the universe, it could in
principle be seen in the CMB and large-scale structure
observations as an increase of the acceleration rate, to be
properly identified as dark energy with variable\footnote{Some
aspects of possible misidentification of varying $w$ as a dark
energy with $w<-1$ have been considered in \cite{maor}.}  $w >
-1$. However those observations do not yet have the ability to
discern such fine structure in $w$ \cite{cmblss}. Hence, the
upward mobility of the quintessence can trick an observer into
deducing $w<-1$ from the geometry of the universe. Nevertheless,
this is accomplished without negative energies or negative norm
states. The effective field theory is perfectly normal.

For such dynamics one needs a  potential for the quintessence
field with a very gentle slope, which the quintessence field has
been climbing for the significant part of the last 14 billion
years, which was preceded by a more curvaceous part. Clearly, one
needs to arrange for the right properties of the potential: the
value of the quintessence potential $V \sim 10^{-12} {\rm eV}^4$,
its curvature ${\partial^2_\phi V} \ga H^2_0$ which changes with
$\phi$, quickly vanishing as $\phi$ exceeds some value ${\mathcal
O}(M_{Pl})$, and sufficiently weak couplings to the visible sector
matter. Of course, this means that the quintessence sector needs
to be finely tuned. The fine tunings are more severe than for the
natural quintessence candidates based on pseudo-scalar Goldstones
\cite{axq}, but not worse than many other oft-quoted quintessence
models in the literature \cite{linq,q,carroll}. On the other hand,
the advent of the Landscape paradigm \cite{landscape} may offer
new means to accomplish the dramatic separation of scales needed
here\footnote{Other interesting cosmological consequences have
been recently discussed in \cite{fkrs}.}. Indeed, a quintessence
field might be some modulus-like degree of freedom, whose
potential arises from SUSY breaking in some hidden sector, which
is transmitted to the quintessence sector very weakly. Typical
such potentials are of the form \cite{banks}
\be V(\phi) = \lambda \, M^4_{Pl} \,  f\left(\frac{\phi}{M_{Pl}}\right)
\, ,
\label{potmod} \ee
where $\lambda$ needs to be tuned to $10^{-123}$, but the
coefficients in the Taylor expansion of $f(x)$ may be numbers of
${\cal O}(10)$. From this point of view the fine tunings do
not look too severe. Such potentials with some mild tuning, so that the
ratio of the slopes on the two sides is of ${\cal O}(10)$, could
perform our task. We
will illustrate the conceptual aspects of the mechanism with a
detailed analysis of the simplest imaginable quintessence field,
which in the region where the field is moving uphill is
approximated by a linear potential
\cite{linq}, \be V(\phi)= \mu^3 \phi~, \label{pots} \ee
where $\mu \sim {\rm few} \times 10^{-13}$ eV, and $\phi({\rm
now}) \sim 10^{19}$ GeV. With a shift of $\phi({\rm now})$, this
is just the leading Taylor series expansion of any quintessence
potential, valid for small enough $\Delta \phi$. Our results can
readily be extended to more complicated potentials. As a result of
running up the potential, the cosmic acceleration is itself
accelerated: it is greater at late times and lower redshifts. This
raises the Hubble diagram at lower redshifts, and in fact is a
slightly better fit to the existing supernova sample \cite{Riess}.
An observer who tries to fit it with quintessence with a constant
$w$ would conclude that a value of $w <-1$ is needed \cite{maor}.
Further, this behavior happens at low redshifts, which is curious
since if there is any support for $w<-1$ in the data, it seems to
arise precisely from such a regime
\cite{starobinsky,huterer,bergstrom}. This is the regime with
relatively few data points, and thus the future searches for the
supernovae in the low redshift range, $0.1 < z < 0.5$ may be
helpful.

Let us now consider the details of the mechanism. In line with the
observations \cite{cmblss}, we will ignore the curvature of the
spatial sections and imagine that the universe is described by a
spatially flat FRW line element,
\be ds^2 = -dt^2 + a^2(t) \, d\vec x^2 \, , \label{metric} \ee
where the scale factor is a solution of the Friedmann equation
\be 3H^2 = \frac{\rho}{M^2_{Pl}} \, . \label{friedman} \ee
The Hubble parameter is $H = \dot a/a$, where the dot represents
the derivative with  respect to the comoving time $t$. We use the
notation where the Newton's constant and the Planck mass are
related by $8 \pi G_N = 1/M^2_{Pl}$. At the scales where the
observations relevant for exploring dark energy are concerned, the
total energy density receives contributions from dark matter and
dark energy. Normalizing the scale factor today to unity, we can
write $\rho$ as
\be \rho = \rho_{cr} \, \frac{\Omega_{M}}{a^3} + \rho_{DE} \, ,
\label{rho} \ee
where $\rho_{cr} = 3 H^2_0 M^2_{Pl}$ is the critical energy
density of the universe now. As we have said above, the dark
energy sector is a quintessence field, with energy density
\be \rho_{DE} = \frac12 \dot \phi^2 + V \, , \label{Qdens} \ee
and whose evolution is controlled by the usual zero mode field
equation
\be \ddot \phi + 3 H \dot \phi = - \, \partial_\phi V \, .
\label{fieldeqs} \ee

As we have explained above, to ensure that the quintessence field
is arrested by friction in the early universe, the potential must
have a mass $m(\phi) = (\partial_\phi^2 V)^{1/2}$ that initially
obeys $m(\phi) < H$. Here $H$ is on the order of the current
expansion rate of the universe $H_0$. Once $H$ is small enough,
but still during the matter domination era, the friction term
looses to the restoring force pulling the quintessence field
toward the minimum. The field begins to roll down the steeper
section (see Fig.~\ref{fig:potential}) of the potential $\sim
m^2(\phi) \phi^2$, picking up speed. Once it enters the gentler
regime of the potential $V(\phi)$, it begins to dominate the
cosmic expansion, but continues to slowly climb the slope since it
still has $\dot \phi_0\ne0$.

To explore the climbing phase for concreteness we will consider a
potential of the form (\ref{pots}) for $\phi>\phi_*$. Here
$\phi_*$ is the initial value of the field at the onset of the
linear potential  regime. We will take $z_*=1$ or $2$ for the
purposes of our calculations. Let us first estimate the values of
the parameters in the above potential that could plausibly
describe our Universe. Since we do not want the kinetic and
potential energies in the scalar field to overclose the Universe,
but still be  non-negligible components of the total energy
density, we have
\ba
&& \frac{1}{2} \dot{\phi}_*^2 \la M_{Pl}^2 H_0^2 \\
&& \mu^3 \phi_* \la M_{Pl}^2 H_0^2 \ea
Finally, we want the variation of the scalar field to be sizable
over a Hubble time, so we require
\be \dot{\phi}_* /H_0 \sim \phi_*. \ee
 From these we see that we get the right orders of magnitudes for
\ba \mu^3 &\la& M_{Pl} H_0^2 \, , \nonumber \\
\phi_* &\ga& M_{Pl} \, , \label{mu} \ea
which is precisely consistent with our choice of the potential
(\ref{pots}). These choices of parameters are also
consistent with (\ref{potmod}), where the quintessence field has a
$4D$ effective theory {\it vev} $\sim {\cal O}(M_{Pl})$. Once the
main fine tuning $\lambda \sim 10^{-123}$ is done, the rest of the
required shape of the potential can be accomplished with mild
tunings of the expansion of $f(x)$.

To gain some more insight into how this system could fake $w<-1$,
let us review the simplest limit when the friction term in
(\ref{fieldeqs}) is negligible. While this does not apply to
the relevant solutions which we have obtained numerically, it is
very useful for illustrational purposes. In this case the field
equation (\ref{fieldeqs}) corresponds to a conservative mechanical
system, and we would find (assuming that the scalar field
stops rolling just about now)
\be
\frac{1}{2}\dot{\phi}_*^2 +V(\phi_*)=V_0.
\ee
Thus we can see that $V_0>V(\phi_*)$, due to the kinetic energy
being converted into potential energy (in general friction
only reduces the amount of kinetic energy which is converted into
the potential energy). An observer who is not aware of the
presence of the kinetic term would interpret this as a growth
of the dark energy density, and infer from it that $w<-1$.
Indeed we will show later using the numerical solutions of the
full equations that the solution for the luminosity distance
vs. redshift curves for the upward rolling scalar field
cannot  be distinguished from that of a phantom matter  at low
redshifts $z<1$. This behavior is completely general for an
arbitrary modulus-like potential (\ref{potmod}) which curves
upward, and not only the linear potential which we have focused
on.

We stress yet again that the true $w$ remains safely above $-1$
throughout the regime we describe, since there are no negative
energy or negative norm states here. Clearly, in a generic case
the slowdown of the field is faster in steeper potentials, and so
the epoch imitating $w<-1$ will be confined to a narrower range of
redshifts.  We will extract quantitative statement about just how
negative the fake $w$ may seem from the numerical fits to the
supernova data of \cite{Riess}.

To get a precise description of the effect we now turn to
numerical integration of the field equations and the explicit
determination of the luminosity-distance relationship. We use the
technique of recasting the field equations in terms of the
redshift instead of the time variable. With our normalizations, $z
= 1/a(t) - 1$, and so the field equations are~\cite{starobinsky}
\ba \frac{\mu^3 \phi (z)}{\rho_c} =\frac{H^2(z)}{H_0}-\frac{1+z}{6
H_0^2}\frac{d H^2(z)}{dz}
-\frac{1}{2} \Omega_M (1+z)^3  \\
\frac{1}{\rho_c} \left( \frac{d \phi (z)}{dz}\right)^2
=\frac{2}{3H_0^2 (1+z)} \frac{d \ln H(z)}{dz} -\frac{\Omega_M
(1+z)}{H^2(z)}. \ea
while the luminosity-distance and the magnitude as a function of
the redshift is given by
\ba
&& D_L(z)=(1+z)H_0 \int_0^z \frac{1}{H(z')}dz' \ , \\
&& m (z)=5 \log_{10} D_L(z).
\ea
To compare with the gold data set of type Ia SNe \cite{sne}, we
first subtract the magnitude in a universe with the best fit
cosmological constant ($\Omega_\Lambda=0.71$)  to get the
magnitude differences \be \Delta m (z)=m(z)-m_{{\rm best}\,
\Lambda}(z)~. \ee We do the same subtraction for the SNIa gold
data set, and to calibrate the overall magnitude of the supernovae
we subtract the average of these relative magnitudes for the
near-by supernovae ($z<0.1$). The difference obtained this way can
be directly compared to the theoretical expressions of $\Delta
m(z)$ above.

For a fixed value of $\Omega_M$ (assuming a flat Universe,
$\Omega_{tot}=1$) we need to specify the fraction of the kinetic
energy $\Omega_{kin}$ of the scalar field today, and the steepness
of the potential $\mu$. This completely fixes the initial
conditions to be
\ba
&& \phi(0) =\frac{(1-\Omega_{kin}-\Omega_M) \rho_c}{\mu^3}  \\
&& H(0) = H_0.
\ea
We then scan over the parameters $\Omega_{kin}$ and $\mu$ for
fixed values of $\Omega_M$ to find the best fit (lowest $\chi^2$
values).

\begin{figure}[tb]
\begin{center}
\includegraphics[width=0.8\hsize]{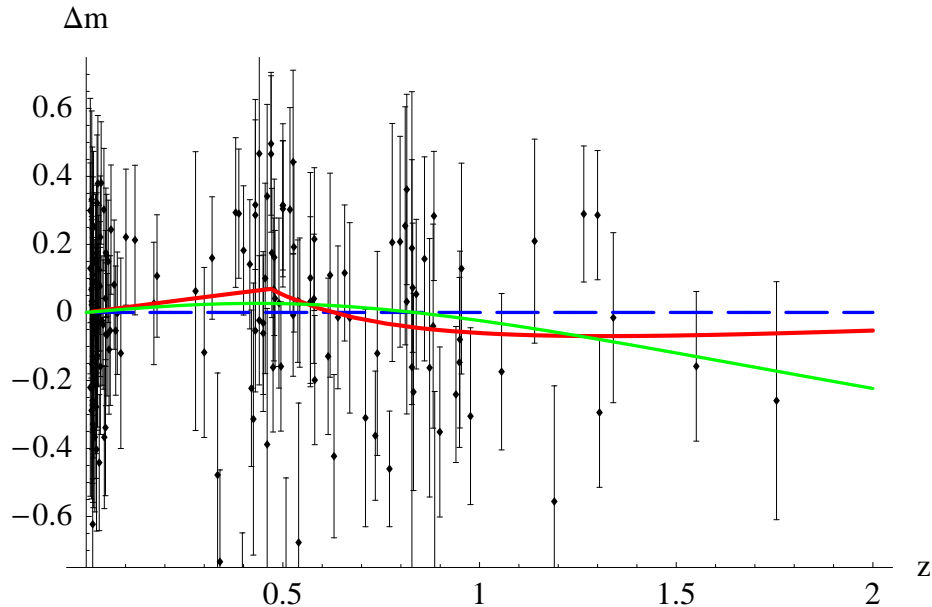}
\end{center}
\caption{Residual magnitudes for high redshift supernovae data
relative to a universe with a best fit
cosmological constant. The dashed line thus gives the prediction
for  $\Omega_\Lambda=0.71$. The green (light) line gives the best
fit for a linear potential with $\Omega_{DE,0}= 0.77$ and $z_*=2$,  
which requires the field to run up the
potential. The red (dark) line gives
the best fit for a change in $w$ from -0.73 for $z>0.47$ to $w=-1$
for $z<0.47$ with $\Omega_{DE}= 0.80$.} \label{fig:residuals}
\end{figure}

In Fig.~\ref{fig:residuals} we have plotted the residual
magnitudes of the type Ia SN gold data sample from \cite{Riess}
(relative to the best fit model with a cosmological constant,
which is given by the flat line). We can see that the experimental
data prefers a slightly more accelerated Universe at low redshifts
(and perhaps a slightly less accelerated  one at high redshifts,
though the data is very sporadic there  and with large error
bars). We have shown two curves in Fig.~\ref{fig:residuals} that
would achieve such modifications in the Hubble plot. The first is
the linear potential model with the field rolling up the hill that
we presented in this paper, which has slightly lower $\chi^2$ than
the best cosmological constant fit. The best fit corresponds to
$\mu^3= H_0^2 \times 2.3 \times 10^{18}$ GeV, and $\Omega_{kin,0}=
8 \times 10^{-4}$. To show that the accelerated acceleration is
robust we also show a second parameterization of the Hubble plot
where the equation of state parameter of the dark energy suddenly
changes at some intermediate value $z_c$. This parameterization
actually gives -- to our knowledge -- the lowest $\chi^2$ of all
the parameterizations tried so far, if the value of $w$ changes
from -0.73 for $z>0.47$ to $w=-1$ for $z<0.47$ with $\Omega_{DE,0}=
0.80$. This jump in $w$ also gives an accelerated acceleration.
However we do not know of a simple physical model that could give
this kind of Hubble diagram. Note, that a step potential would not
be equivalent to this parameterization, but would rather give  a
curve very similar to the linear potential case.

\begin{table}[htb]
\be
\begin{tabular}{c|c|c|c|c|c|c}
model & $\Omega_{DE,*}$ & $\Omega_{DE,0}$ & $\Omega_{M,*}$
& $\Omega_{M,0}$ & $\chi^2$  &
$\chi^2/$(d.o.f.) \\
\hline
cosm. const. & 0.08 & 0.71 &0.92 & 0.29 & 177.4 & 1.14 \\
$w = -1.2$ & 0.03 & 0.65 & 0.97 & 0.35 & 176.0 & 1.14 \\
linear pot. & 0.02 & 0.74 & 0.36& 0.26 & 175.8 & 1.14 \\
linear pot. & 0.01 & 0.77 & 0.26& 0.23 & 175.6 & 1.14 \\
$w = -1.4$ & 0.01 & 0.6 & 0.99 & 0.4 & 175.0 & 1.13 \\
$w$ jump & 0.21  & 0.80 & 0.79 &0.20 & 172.0 & 1.12
\end{tabular} \nonumber
\ee \caption{\label{table} The best $\chi^2$ fits for different
dark energy models for different values of $\Omega_{M,0}$.
The subscript $*$ refers to $z_*=2$. The second and fifth
rows show the best fits for  phantoms, while the last row shows a
hypothetical model where the equation of state of the dark energy
suddenly jumps (as explained in the text).}
\end{table}

In Table~\ref{table} we present a summary of the fits to different
dark energy models.  We include four models: a cosmological
constant, the linear potential model elaborated on in this paper,
a model with phantom matter, and a hypothetical model with a jump
in the equation of state as explained above. We can see that the
minimal $\chi^2$ of the linear potential model is lower than that
for the cosmological constant (and comparable to that of the
phantom), however since there is one more parameter in this model
the $\chi^2/(\rm d.o.f.)$ is comparable to the case of the
cosmological constant. The best fit is for the model with a
jump in $w$. However, of the scenarios with a scalar in a
potential the best fit is always for a field
rolling up a potential.  It is interesting
to note that the accelerated acceleration model prefers a low
value of $\Omega_M$ while the phantom prefers a large value of
$\Omega_M$, however the accelerated acceleration model is less sensitive
to the value of $\Omega_M$. Thus an improved determination of  
$\Omega_M$ would
help to distinguish between the two.

\begin{figure}[htb]
\begin{center}
\includegraphics[width=0.4\hsize]{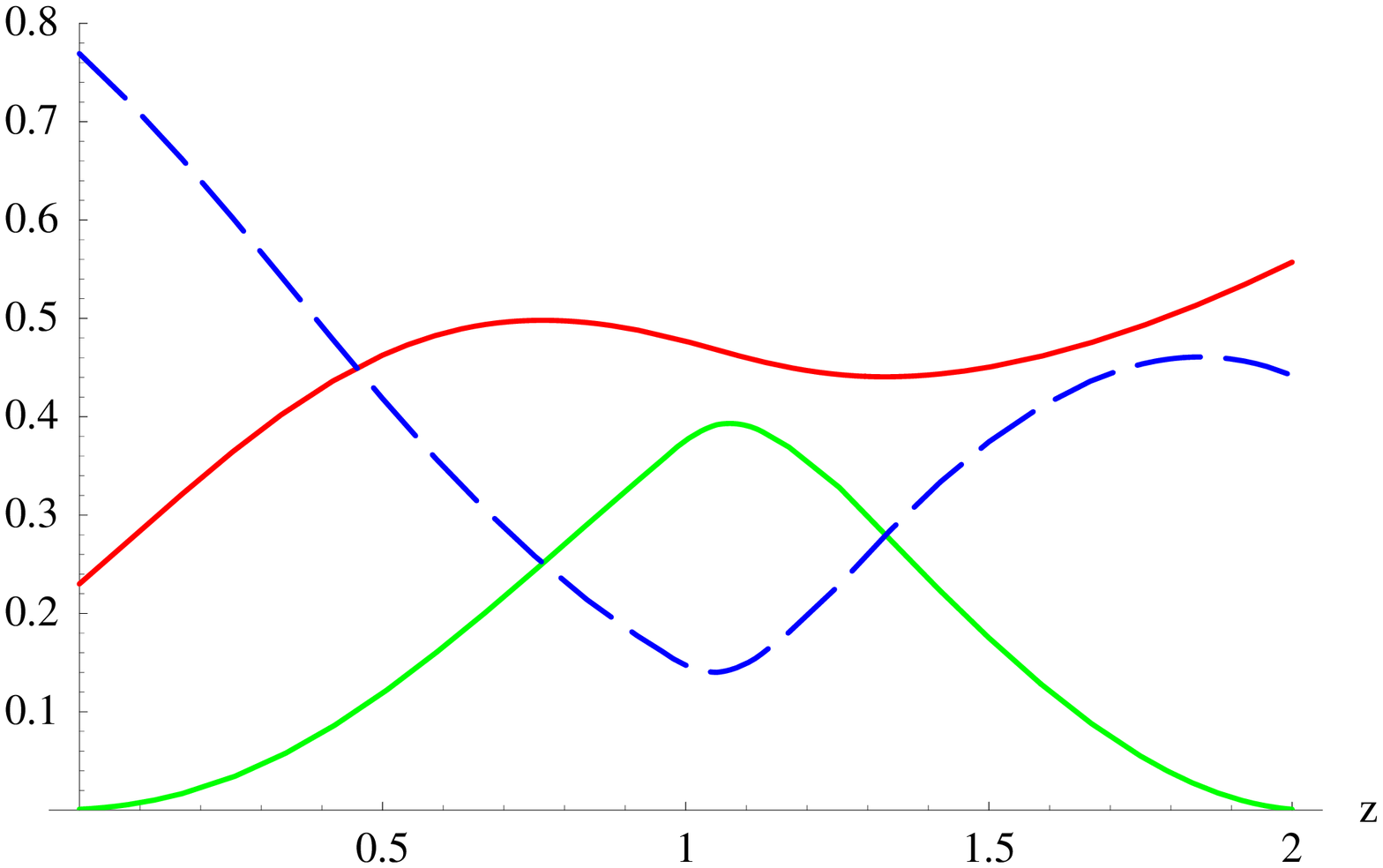}\includegraphics[width=0.4\hsize 
]{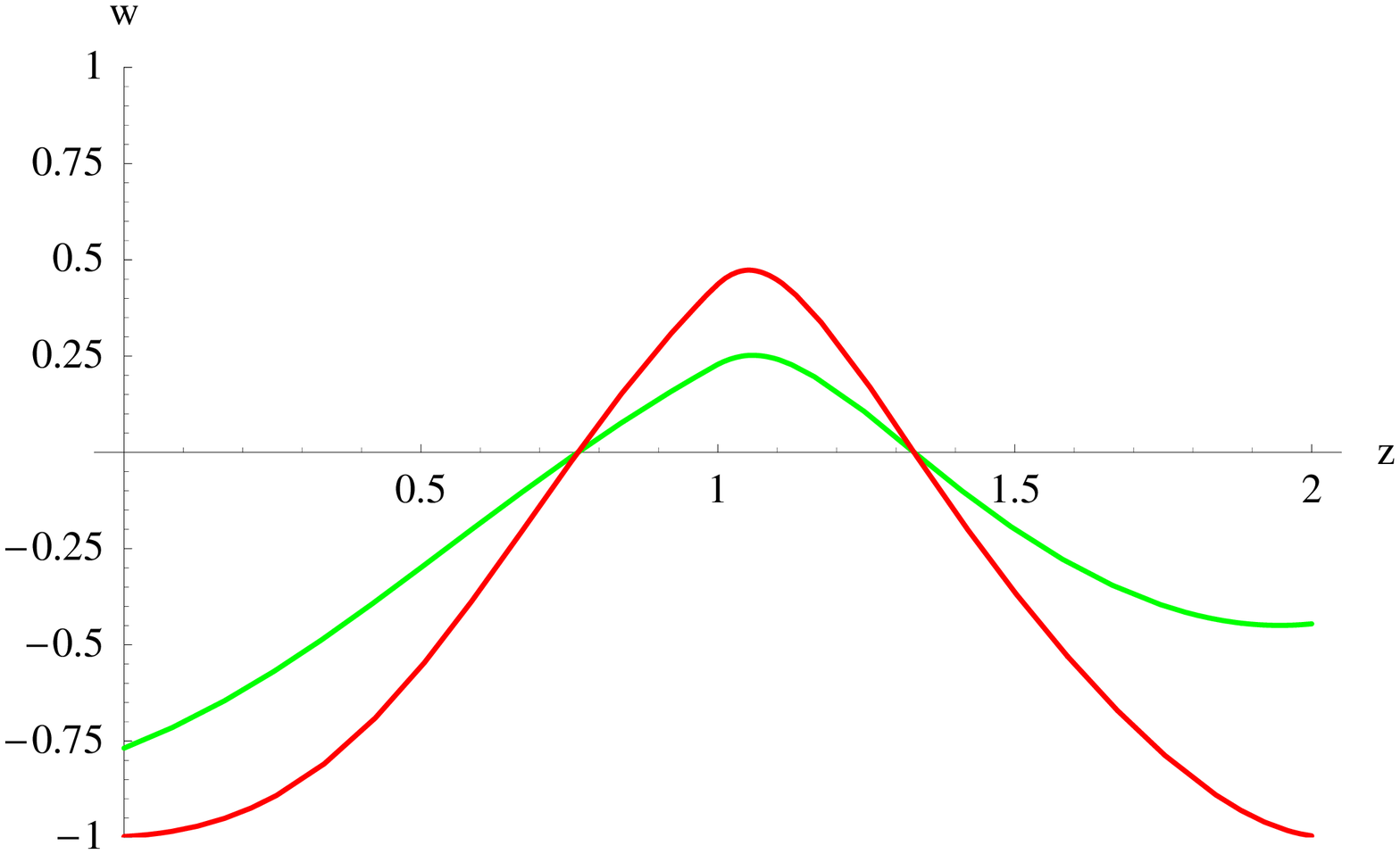}
\end{center}
\caption{On the left we show $\Omega_M$ as a function of $z$ with
the red (dark) line which dominated for $z>0.5$, $\Omega_{kin}$ with the  
green (light) line,
and $\Omega_{DE}$ with the dashed line for a scalar field with the best
fit
linear potential for $z<1$ and with a simple toy model for the  
asymmetric
part of the potential for $z>1$ as described in the text. On the right  
we
show
in
red (dark) the time dependent effective equation of state
parameter $w_{eff}$ for the scalar field in the same scenario,
while in green (light) we
show the full $w_{eff}$ including dark matter. } \label{fig:w}
\end{figure}

In Fig.~\ref{fig:w} we show the energy density components and the
effective equation of state parameter for the best fit linear
potential case. The definition  of  $w$ for a scalar field is
found by comparing the $00$ and $ii$ terms in the energy-momentum
tensor \be w_{eff}=\frac{\frac{1}{2}\dot{\phi}^2-V(\phi
)}{\frac{1}{2}\dot{\phi}^2+V(\phi )}. \ee To keep $\Omega_{kin}$
from dominating the universe  we have put in a simple toy model
for the behavior of the asymmetric part of the potential. We
matched onto a quadratic plus linear potential that makes the
potential and its first derivative continuous at $\phi(z_*=1)$. Here we
have chosen the parameters of the potential (for $z>1$) so as to
avoid kinetic energy domination at all times.  The behavior for
$z<1$ is well constrained by the supernova data, but is quite
model dependent for $z>1$.
Note that the modification of $w_{eff}$ at redshifts $1<z<3$ could
affect structure but only at the largest scales, and in a model
dependent way. More detailed work would be needed to evaluate
these effects.

\begin{figure}[tb]
\begin{center}
\includegraphics[width=0.8\hsize]{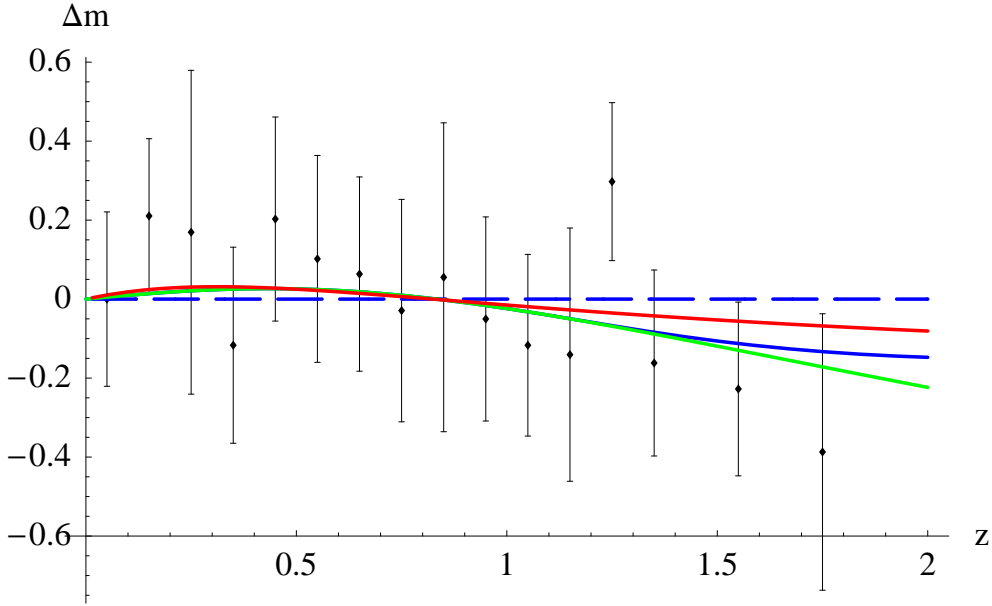}
\end{center}
\caption{Residual magnitudes relative to a flat universe with a
best fit cosmological constant. Data points are SN Ia gold data in
redshift bins of 0.1. The bottom, green (light) line gives the best fit
for a linear potential,  with $\Omega_{DE,0}= 0.77$.  The top, red
(dark) line gives the best fit for a phantom with  $w=-1.4$  with
$\Omega_{DE}= 0.6$. The blue (darkest) line gives the behavior when the
linear potential
is matched onto a quadratic potential at $\phi(z_*=1)$, as described in  
the
text.} \label{fig:fakew}
\end{figure}

In Fig.~\ref{fig:fakew} we show the Hubble plots for the model
with the best fit linear potential with $z_*=2$, the linear potential  
matched to a quadratic at $\phi(z_*=1)$,
  the best fit phantom,
  and the binned supernova data. It is clear from the binned
data why an accelerated acceleration is preferred for the fit. All
but one of the the data points that lie above the cosmological
constant  fit  occur with $z<0.9$ while all but one of the data
points that lie below the cosmological constant fit occur for
$z>0.9$.  This seems to show a systematic trend. We can see that
for $z<1$ the curves for the phantom and the linear potential are
practically indistinguishable from each other, and hence the
linear potential model could easily be mistaken as a Universe with
phantom matter. For larger values of $z$ the two curves start
deviating from each other. A precision measurement of SN Ia from
future dark energy probes may be able to distinguish between these
two models, however the region $1<z<2$ is a more model dependent
part of this scenario because it depends on the details of the
non-linear part of the potential.

To conclude, we have presented a very simple model of a
conventional quintessence field which can mimic the equation of
state $w<-1$ without any ghosts, negative energies or negative
norm states. The key ingredient is to imagine that during the
epoch of current cosmic acceleration the quintessence field has
been persistently moving up the potential slope. This requires
fine tunings, but of a kind familiar from a generic quintessence
setup. The advantage of the mechanism is that it provides a
simple, yet potent method of impersonating $w<-1$ in a way which
is completely grounded in conventional $4D$ physics. At the
moment, this behavior provides a lower $\chi^2$ fit to the SN data
than a genuine cosmological constant. Searches for more supernovae
in the regime of  low redshifts $0.1 < z  < 0.5$, where the
turnover of the luminosity distance-redshift curve is located, and
at high redshifts $z>1$, where the luminosity distance-redshift
curve dips below the one for cosmological constant will improve
the precision for the fits and may be instrumental for discerning
accelerated acceleration from a cosmological constant.

\section*{Note added}
While this manuscript was in preparation Ref.~\cite{Liddle}
appeared which also considers models with fields rolling up a
linear potential. They note that these models would in fact give
the best fits to the data, however discard this scenario due to
the assumption that the potential is linear for all values of
$\phi$. We have argued here that an asymmetric potential could
easily remove the unwanted kinetic energy dominated phase for
large $z$ without changing the analysis for $z<2$.

\section*{Acknowledgments}

We thank A. Cohen, T.~Damour, J.~Frieman, M.~Kaplinghat,
M.~S.~Sloth and L.~Sorbo for useful discussions. C.C. and J.T. are
grateful to the Aspen Center for Physics for kind hospitality
during the completion of this work. C.C. thanks the particle theory
group at UC Davis for kind hospitality during the early stages of this
work. N.K. is grateful to the {
Institut des Hautes Etudes Scientifiques}, Bures-sur-Yvette,
France, for kind hospitality during the completion of this work.
The work of C.C. is supported in part by the DOE OJI grant
DE-FG02-01ER41206 and in part by the NSF grants PHY-0139738 and
PHY-0098631. The work of N.K. is supported in part by the DOE Grant
DE-FG03-91ER40674, in part by the NSF Grant PHY-0332258 and in
part by a Research Innovation Award from the Research Corporation.


\end{document}